# Effect of Anisotropy on Defect Mode Peculiaries in Chiral Liquid Crystals

A.H. Gevorgyan[1,2], K. B. Oganesyan[3,4]

[1]Yerevan State University, 1 Al. Manookian St., 025, Yerevan,Armenia

[2]Ins. of Applied Problems in Physics, 26, Hr. Nersessian, 0014, Yerevan, Armenia

[3]Alikhanyan National Science Lab, Yerevan Physics Institute, Alikhanyan Br.2, 036, Yerevan, Armenia, bsk@.yerphi.am

[4]Laboratory of Information Technologies, JINR, Dubna, Russia

## Abstract

 The effect of anisotropy on defect mode peculiarities in cholesteric liquid crystals is investigated. The problem is solved by Ambartsumian's layer addition modified method. Two cases are considered. In the first case, it is assumed that the defect layer is not absorbing and the effect of refraction anisotropy on the reflection, relative  photonic density of states and the total field intensity aroused in the defect layer are  studied .  In the second case, defect layer is assumed to be isotropic for refraction and anisotropic for absorption, and the influence of defect layer absorption anisotropy on reflection, absorption, relative photonic density of states and the total field intensity aroused in the defect layer are investigated.

## 1. Introduction

In the last three decades, micro-nano-structuring is a very promising method for controlling the optical properties of materials. This method reflects a new approach of modern optics, in which optical properties are mainly determined not by electronic or magnetic resonances, but by geometric factors.

An example of the successful implementation of this approach are photonic crystals (PC). PC are one of the most interesting structures in optics, which make it possible to form the reflection and transmission zones of electromagnetic radiation. The main theme that attracts increasing attention is the active adjustment of the width of the photonic band gap (PBG) and its frequency location. PCs have many applications, and devices based on PC have a number of advantages. However, the presence of a defect in the structure of the PC makes them more useful, like doped semiconductors. Introducing a defect into the PC structure leads to additional resonance modes inside the PBG. Such defect modes are localized in the defect positions and can be used for constructing narrow band filters and mirrors.

PBG and defective modes also significantly change the distribution of the photonic density of states (PDS) of the material. This effect was widely used to change spontaneous and stimulated emission processes that strongly depend on the interaction between the emitter and its local electromagnetic environment.

In various optoelectronic devices, such as light-emitting diodes, solar cells, optical amplifiers and low-threshold lasers, the effects of PBG and defect modes make it possible to improve overall efficiency. PCs with easy tunable parameters are of great interest. The well-known representatives of soft PCs are cholesteric liquid crystals (CLCs) and blue phases of liquid crystals. CLC layers as PCs have polarizations peculiarities and their defect modes possess polarization peculiarities, too [1-13]. Recently there have been many activities in the field of localized optical modes, edge modes and defect modes in CLCs due to the possibilities to reach a lasing in CLCs. In the general case, lasers on the base of CLC layers can be separated into three classes: (1) lasers working on edge modes, (2) lasers on defect modes, and (3) random lasers. Cholesteric LCs with defects of different types have been studied in the last decades to gain insight into the specific features of lasing on defect modes, characteristics of absorption/emission on these modes, etc. (see [1–13] and references therein) (for integrated photonics see [19,20] ).

In this paper we will investigate the effect of optical anisotropy of a defect layer on the features of defect modes and the optical properties of a CLC with a defect inside (Fig. 1). A CLC layer with a defect inside with a controlled anisotropy can be created by introducing a layer of a nematic liquid crystal with the possibility of electrically controlling the orientation of its optical axis or simply allowing local control of the direction of the optical axis of the CLC (Fig. 1). A change in the angle $\varphi$ by means of an external static electric field leads to a change in the anisotropy of the defect layer $\Delta = \dfrac{\varepsilon_x - \varepsilon_y}{2}$ where $\varepsilon_x$ and $\varepsilon_y$ are x and y components of dielectric permittivity.

## 2. Results and discussion

The problem is solved by Ambartsumian's modified layer addition method adjusted to the solution of such problems (see [6]). A CLC layer with a defect layer inside (with a DL) can be treated as a multilayer system: CLC(1) − (DL) − CLC(2) (Figure 1). The ordinary and extraordinary refractive indices of the CLC layers are taken to be $n_o = \varepsilon_2 = 1.4639$ and $n_e = \varepsilon_1 = 1.5133$; $\varepsilon_1$, $\varepsilon_2$ are the main values of the CLC local dielectric tensor. The CLC layer helix is right handed and its pitch is: $p = 420$ nm. The parameters of the CLC cholesteryl − nonanoate − cholesteryl − chloride −cholesterylacetate are (20 : 15 : 6) composition at the temperature $t = 25$ °C. Hence, the light normally incident onto a single CLC layer − with right circular polarization (RCP) − has a PBG (which is in the range of $\lambda \sim$ (614.8–635.6) nm), and the light with left circular polarization (LCP) does not have any. We consider the case when our system is surrounded by the media with the reflection index $n_s$ on its both sides, which is equal to the CLC average reflection index $n_m = \sqrt{\varepsilon_m} = \sqrt{\dfrac{\varepsilon_1 + \varepsilon_2}{2}}$. CLC layer thickness is $d$=60$p$ and defect layer thickness is $d^d$=150nm.

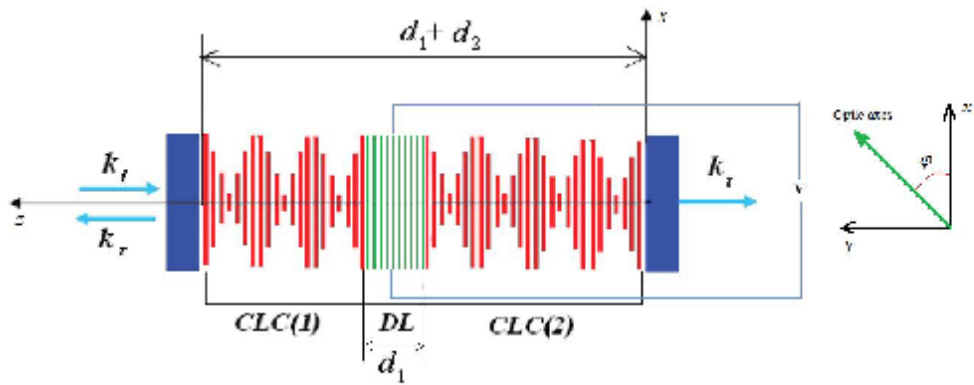

**Figure 1.** A sketch diagram of a modeled CLC cell with a defect layer inside.

Representing the main values of the permittivity tensor of the defect layer in the form $\varepsilon_{1,2}^d = \varepsilon_m \pm \Delta$, we will investigate the effect of the anisotropy of $\Delta$ on the reflection $R$, the relative PDS $\rho / \rho_{iso}$ and the total field intensity $I = |E|^2$ aroused at the left border of the defect layer.

The PDS, i.e. the number of wave vectors $k$ per unit frequency $- \rho(\omega) = dk / d\omega -$ is the reverse of the group velocity and can be defined by the expression [14]:

$$\rho_i(\omega) \equiv \frac{dk_i}{d\omega} = \frac{1}{d} \frac{\dfrac{du_i}{d\omega} v_i - \dfrac{dv_i}{d\omega} u_i}{u_i^2 + v_i^2}, \; i = 1, 2 \tag{1}$$

where $d$ is the whole system thickness, $\omega$ is the incident light frequency, and $u_i$ and $v_i$ are the real and imaginary parts of the transmission amplitudes; $t_i(\omega) = u_i(\omega) + jv_i(\omega)$ are the transmission amplitudes for the incident light with the two eigen polarizations (EPs), $j$ is the imaginary unit. The EPs are the two polarizations of incident light, which do not change when light transmits through the system. For the single CLC layer the two EPs practically coincide with the two circular (right and left) polarizations. The values $i=1$, 2 correspond to the diffracting and non-diffracting EPs, respectively. For a CLC layer with a defect layer inside a jump of EPs takes place at the defect mode. For the isotropic case we have: $\rho_{iso} = n_s / c$, and $c$ is the speed of light in vacuum.

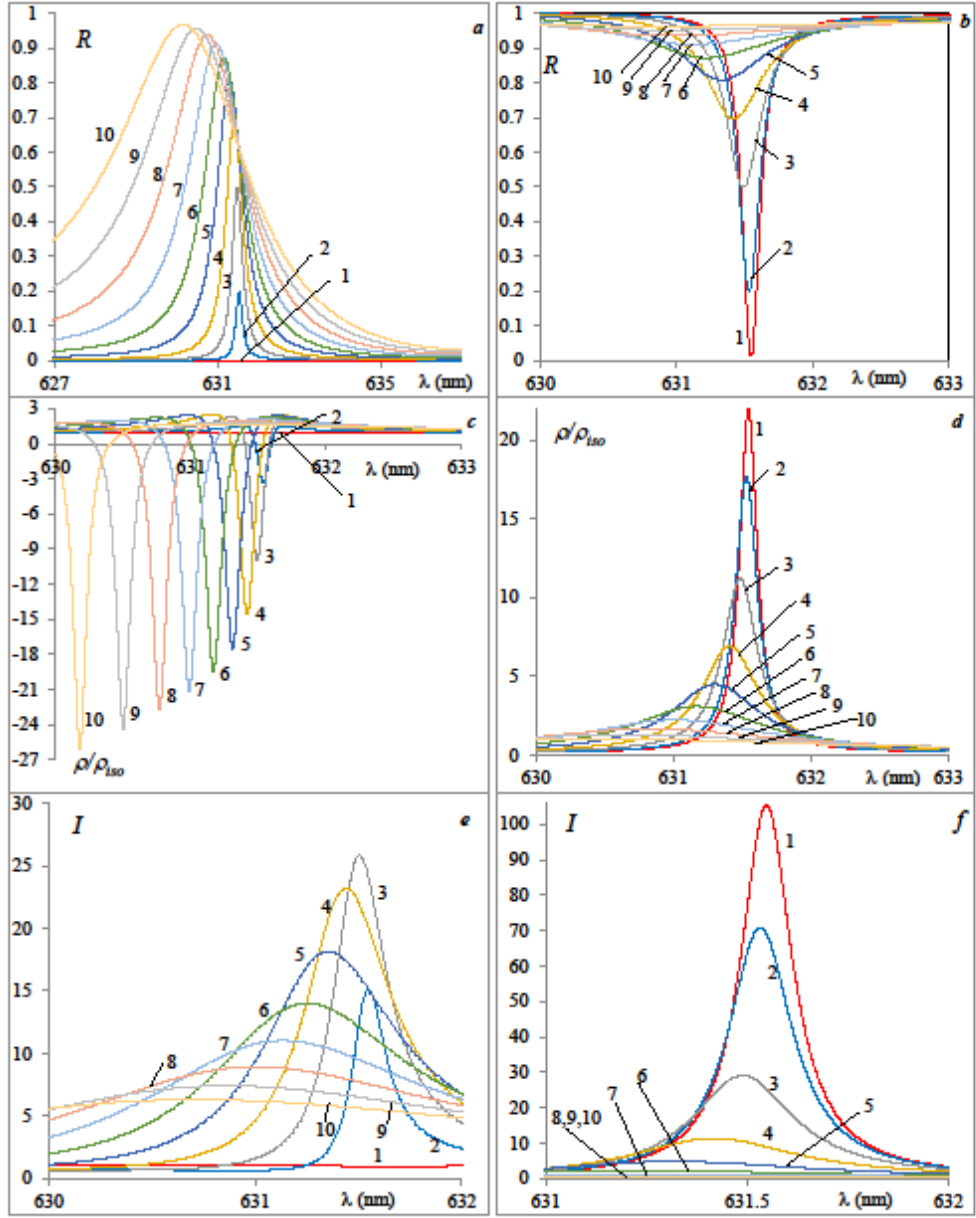

**Figure 2.** (*a, b*) the reflectance *R,* (*c, d*) relative PDS $\rho/\rho_{iso}$ and (*e, t*) the total field intensity $I=\left|E\right|^2$ spectra at different values of anisotropy of the defect layer $\Delta$ ($\Delta=0.1(i$-1), where $i$ is the number of curve). The incident light on the layer has a left (left column) and right (right column) circular polarizations.

Figure 2. shows (*a, b*) the reflectance *R,* (*c, d*) relative PDS $\rho/\rho_{iso}$ and (*e, t*) the total field intensity $I=\left|E\right|^2$ spectra at different values of anisotropy of the defect layer $\Delta$ ($\Delta=0.1(i$-1), where $i$ is the number of curve). The incident light on the layer has a left- (left column) and right- (right column) circular polarizations.

As is well known, (see, for example [6]), the defect modes aroused in CLC are either narrow transmittance lines in PBG for diffracting circular polarization of the incident light, or they are narrow lines of reflection in the transmittance band for the not diffracting circular polarization of the incident light.

At $\Delta = 0$, the defect mode for incident light with nondiffractive circular polarization under the condition $n_o^d = n_e^d = n_s = n_m$ is absent [15]. The dependence of the reflection on polarization explains the appearance of a peak in the reflection spectrum for light with left circular polarization (nondiffractive circular polarization). Under these conditions, Fresnel - type reflections at the boundaries of the defect layer are absent.

With increasing $\Delta$, a peak appears in the reflection spectrum for light with left circular polarization, and with a further increase in the anisotropy, reflections at the peak increase, the width of the defect mode line increases, and the defect mode shifts toward the short-wave boundary of the PBG.

As was mentioned above, the presence of a defective layer in the CLC structure leads to the appearance of a dip in the reflection spectrum in the PBG for light from the right (diffracting) circular polarization. A narrow transparency window appears in the PBG. With increasing $\Delta$, the reflections decrease on the defect mode dip, the width of the defect mode line increases, and the defect mode shifts toward the short-wave boundary of the PBG.

The sizes for the displacement of defect modes for right and left - handed polarized incident waves are not equal. The regularities in the spectra (Figures 2 *c, d*) are also easy to understand. The following should be noted concerning negative values of $\rho / \rho_{iso}$ . One must realize that the concept of the PDS introduced by equation (1) is not without controversy, especially in the case of non-periodic systems (for more detail about this see in [16]). Let us only note that although it is impossible to ascribe a direct physical meaning to the PDS in equation (1) in a general case, anyhow, it can be used as a parameter, which can provide some heuristic guidance in experiments for the PDS dispersion - related effects. Moreover, as was shown in [6–25], equation (1) is applicable for finite PCs layers, even for cases with absorption and amplification (naturally, for weak ones).

As is known, a strong accumulation of light takes place in a defective mode [12]. Dependence $I = \left| E \right|^2$ on the wavelength for incident light with diffractive circular polarization has a resonant form. With increasing of $\Delta$, the accumulated light energy decreases on the defect mode, the peak shifts toward short waves and at the same time the line broadens.

At $\Delta = 0$ for incident light with non-diffracting circular polarization, under the condition $n_o^d = n_e^d = n_s = n_m$ the accumulation of light energy on the defect does not occur. With increasing $\Delta$, a peak appears in the spectrum of $I = |E|^2$, and with increasing of anisotropy, the height of this peak first increases, and then passing through the maximum, the height of this peak begins to decrease, the line width $I(\lambda)$ increases simultaneously and the peak shifts toward short waves.

The evolution of the total field intensity $\ln(I) = \ln\left(|E|^2\right)$ spectra when the optical anisotropy of defect layer $\Delta$ increases is presented in Fig. 3. for the $(a)$ non-diffracting and $(b)$ diffracting EPs.

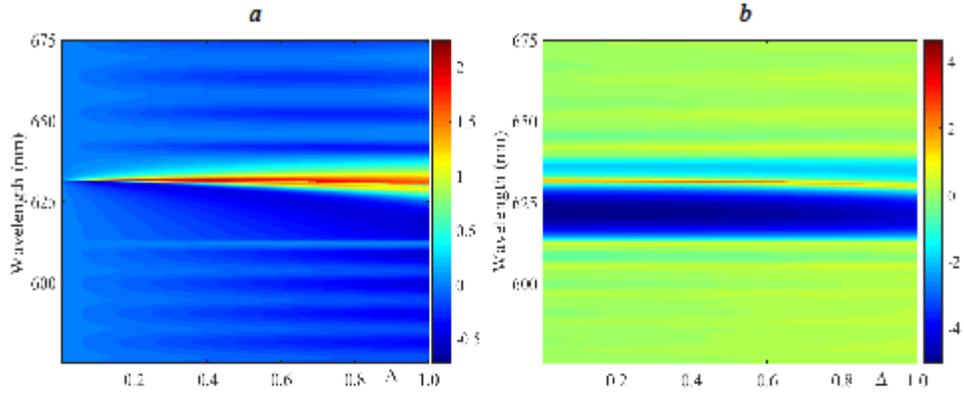

**Figure 3.** The evolution of the total field intensity $I = |E|^2$ spectra when the optical anisotropy of defect layer $\Delta$ increases for the $(a)$ non-diffracting and $(b)$ diffracting EPs.

To get full idea on the effect of each of factors on the properties of the defect modes in CLCs, we will consider here a limiting case, when the defect layer has the property $\mathrm{Re}\,\varepsilon_1^d = \mathrm{Re}\,\varepsilon_2^d$, but $\mathrm{Im}\,\varepsilon_1^d \neq \mathrm{Im}\,\varepsilon_2^d$, where $\varepsilon_{1,2}^d$ are the main values of the dielectric constant tensor of the defect layer.

Representing the imaginary parts of the main values of the dielectric tensor of the defect layer in the form $\mathrm{Im}\,\varepsilon_2^d = 0, \mathrm{Im}\,\varepsilon_2^d = 0.002\,(i-1)$, where i is the number of curve i = 1,2, ... 10, we will investigate the effect of absorption anisotropy on reflections $R$, absorbed in the system light energy $A$ ($A = 1$- $(R + T)$, $R$ and $T$ are the reflection and transmission coefficients, respectively), relative PDS $\rho / \rho_{iso}$ and the total field intensity $I = |E|^2$ aroused at the left border of the defect layer.

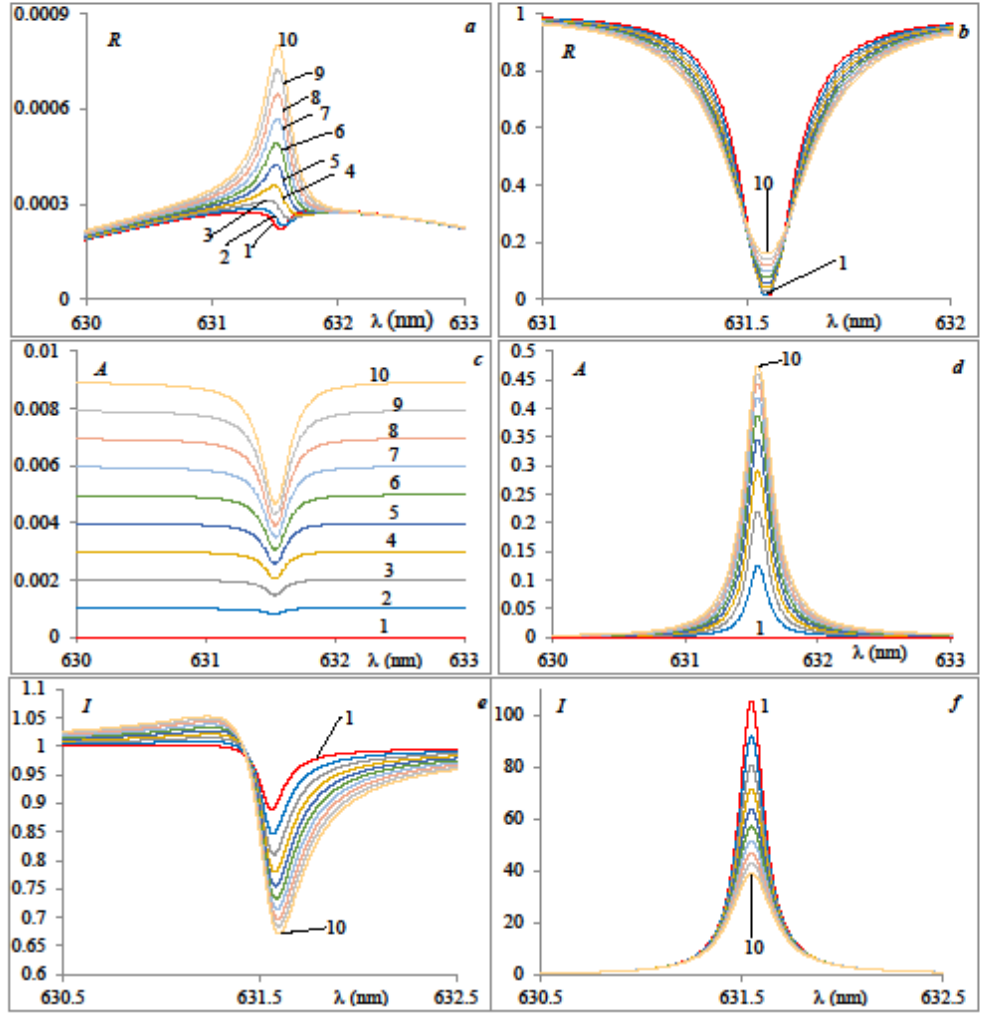

**Figure 4.** (*a, b*) the reflection *R*, (*c, d*) the absorption *A*, (*e, f*) the total field intensity $I = |E|^2$ at the left border of the defect layer and (*g, h*) the relative PDS $\rho / \rho_{iso}$ spectra at different values of $\mathrm{Im}\,\varepsilon_1^d = 0.002\,(i-1)$, where *i* is the number of curve. $\mathrm{Im}\,\varepsilon_2^d = 0$. The incident light on the layer has a left (left column) and right (right column) circular polarizations.

Figure 4 shows (*a, b*) the reflection *R*, (*c, d*) the absorption *A*, (*e, f*) the total field intensity $I = |E|^2$ at the left border of the defect layer and (*g, h*) the relative PDS $\rho / \rho_{iso}$ spectra at different values of $\mathrm{Im}\,\varepsilon_1^d = 0.002\,(i-1)$, where *i* is the number of curve. The light incident on the layer has a left (left column) and right (right column) circular polarizations.

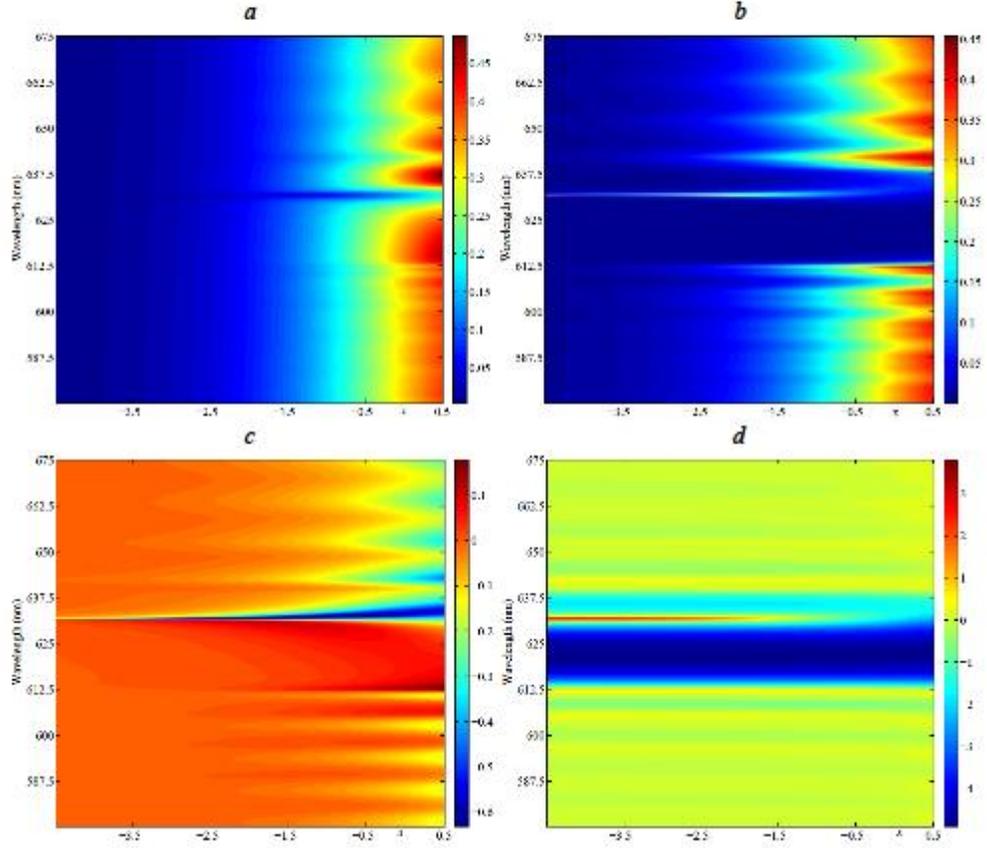

**Figure 5.** The evolution of the $(a,b)$ absorption $A$ and $(c,d)$ total field intensity $I = |E|^2$ spectra when the $x = \operatorname{Im}\varepsilon_1^d$ of defect layer increases for the $(a,c)$ non-diffracting and $(b,d)$ diffracting circular polarizations. $\operatorname{Im}\varepsilon_2^d = 0$.

The increase of $\operatorname{Im}\varepsilon_1^d$ leads to a decrease in the reflection of $R$, to an increase in the absorption $A$, to a decrease in the accumulation of light, and to a decrease in the relative PDS in the defect mode for incident light with diffractive circular polarization. The presence of weak absorption leads to the appearance of a dip in the reflection spectrum on the defect mode for light with non-diffracting circular polarization. The increase of $\operatorname{Im}\varepsilon_1^d$ leads to the replacement of this dip to the peak and the height of this peak increases with a further increase of $\operatorname{Im}\varepsilon_1^d$. The increase of $\operatorname{Im}\varepsilon_1^d$ leads to a decrease in the accumulation of light and hence to the suppression of the absorption $A$ in the defect mode for incident light with non-diffracting circular polarization. The increase of $\operatorname{Im}\varepsilon_1^d$ leads to weak changes in the relative PDS spectrum near the defect mode for incident light with non-diffracting circular polarization.

The evolution of the (*a, b*) absorption *A* and (*c, d*) total field intensity $I = |E|^2$ spectra when the parameter $x = \mathrm{Im}\,\varepsilon_1^d$ of defect layer increases for the (*a, c*) non-diffracting and (*b, d*) diffracting circular polarizations are presented in Fig.5 for the sake of completeness.

### 3. Conclusion

The effect of anisotropy on defect mode peculiarities in CLCs is investigated. Two cases are considered. In the first case, it is considered that the defect layer is not absorbing and has the property of anisotropy of refraction. In the second case, it is considered that the defect layer has the properties of isotropy of refraction and anisotropy of absorption. The influence of the corresponding parameters on reflection, absorption, relative photonic density of states, and the total field intensity aroused in the defect layer are studied.

The effect of suppressing the absorption on the defect mode for incident light with non-diffractive polarization and the anomalous strong absorption effect on the defect mode for incident light with diffractive polarization are discovered.